\documentclass[manuscript]{acmart}
\usepackage{todonotes}
\AtBeginDocument{%
  \providecommand\BibTeX{{%
    \normalfont B\kern-0.5em{\scshape i\kern-0.25em b}\kern-0.8em\TeX}}}

\setcopyright{acmcopyright}
\copyrightyear{2022}
\acmYear{2022}
\acmDOI{XXXXXXX.XXXXXXX}

\acmConference[InContext Workshop at CHI '22]{Make sure to enter the correct
  conference title from your rights confirmation emai}{May 1,
  2022}{New Orleans, LA}
\acmBooktitle{Proceedings of the InContext: Futuring User-Experience Design Tools Workshop at CHI Conference on Human Factors in Computing
Systems (CHI ’22), May 1,
  2022, New Orleans, LA, USA} 
\acmPrice{15.00}

\begin{document}

\title{Integrating Social Media into the Design Process}

\author{Morva Saaty}
\authornote{Both authors contributed equally to this research.}
\email{trovato@corporation.com}
\author{Jaitun V. Patel}
\authornotemark[1]
\affiliation{%
  \institution{Department of Computer Science, Virginia Tech}
  \city{Blacksburg}
  \state{VA}
  \country{USA}
}

\author{Derek Haqq}
\affiliation{%
  \institution{Department of Computer Science, Virginia Tech}
  \city{Blacksburg}
  \country{USA}
}

\author{Timothy L. Stelter}
\affiliation{%
  \institution{Department of Computer Science, Virginia Tech}
  \city{Blacksburg}
  \country{USA}
}

\author{D. Scott McCrickard}
\affiliation{%
  \institution{Department of Computer Science, Virginia Tech}
  \city{Blacksburg}
  \country{USA}}







\renewcommand{\shortauthors}{Saaty and Patel, et al.}

\begin{abstract}
Social media captures examples of people's behaviors, actions, beliefs, and sentiments. As a result, it can be a valuable source of information and inspiration for HCI research and design. Social media technologies can improve, inform, and strengthen insights to better understand and represent user populations. To understand the position of social media research and analysis in the design process, this paper seeks to highlight shortcomings of using traditional research methods (e.g., interviews, focus groups) that ignore or don't adequately reflect relevant social media, and this paper speculates about the importance and benefits of leveraging social media for establishing context in supplement with these methods. We present examples that guide our thinking and introduce discussion around concerns related to using social media data. 
\end{abstract}





\maketitle

\section{Introduction}
Social media has become a ubiquitous part of our daily lives. People use social media to communicate with others beyond geographical borders \cite{bhatti2021parenting}, share information, form networks of relationships, discuss common interests \cite{haqq2021toward}, and express their opinions. Hence, there is a lot of information from various demographics with different perspectives worldwide on social media platforms. Communication has also evolved in multiple ways and formats in the digital age and remote life. With the advancement of technology, people have fewer face-to-face interactions and leverage modern communication methods to stay connected \cite{baruah2012effectiveness, haqq2021toward}. For instance, social media revolutionized how information is disseminated and how successfully people communicate \cite{akram2017study}. Social media platforms are evolving into much more than purely social interaction. Shopping, gaming, advertising products, learning, and live streaming are just a few of the many things individuals can do by using social media \cite{appel2020future, haqq2021toward} that was formerly limited to texting a friend or posting a daily status update.

Analysis of social media has been used in the design process.  Studies in industry and academia are adopting social media to replace or use along with traditional research methods like focus groups, ethnography, and interviews during users' research on the front end of the design lifecycle \cite{lambton2020unplatformed, snelson2016qualitative}. Due to the variety of issues and criticisms of conducting classic research methods (e.g., interviews) to acquire data \cite{turner2010qualitative}, namely time-consuming scheduling, misunderstanding created between interviewers and interviewees, and potential of observer's bias, social media technologies have attracted significant attention among HCI researchers and designers. 

Due to the impacts of social media on human behaviors and social interactions \cite{ruths2014social}, it is becoming an essential venue for research and design interventions. Using existing social media platforms as a design resource has a lot of potential for long-term and scalable coordinating participation \cite{lambton2020unplatformed} and gathering unrestricted expressions and reactions of users. HCI researchers and designers leveraged social media platforms as a rich medium to conduct research, and design practices \cite{altarriba2021social, haqqre}. Prior studies represented the potential of social media in conducting online focus groups \cite{lambton2020unplatformed}, influencing topical, emotional, and rhetorical behavior in a broad audience \cite{rho2020political}, gathering user perceptions and trends \cite{ofli2017saki, deeb2017selfie, kotut2020preparing}, and inspiring ideas for future design \cite{pometko2021drawing}.

In this paper, we propose extending the role of social media as a design tool for contextualizing the experiences and actions of research subjects, beyond just addressing research questions in a standalone way. Analyzing the social media usage of research subjects can better inform the data collection and increase the credibility of traditional research methods (e.g., interviews, focus groups). Social media analysis for qualitative research has been limited to formulating interview questions or the themes of focus groups where the responses of the research subjects are documented unfiltered. While such approaches claim to capture the ground realities and wholesome user experiences, they are completely controlled by the research subjects. They raise questions about the reliability of this information. So, for a qualitative study, we envision social media as a tool to strengthen traditional findings, identify any misrepresentations in user responses, and probe deeper insights. 


\section{Insights}
HCI researchers and practitioners use various research methods (e.g., interviews, focus groups, etc.) to understand users' feelings, ideas, and experiences when using a product \cite{saaty2021study, lazar2017research, blandford2016qualitative, saaty2021exergames, Zimring2014}. They use these methods to gather users' insights in the early stages of the research phase or after implementing the product to investigate users' experiences. The user research and corresponding collected information will inform different stakeholders (e.g., product designers, product developers, UX designers, product managers, etc.) about what users perceive about the product and how they can improve the product in the next versions.

The mentioned practical and traditional research methods (e.g, interviews, focus groups, etc.) used in HCI have some limitations that can make them unreliable. For instance, group dynamics can be skewed by strong opinions in focus groups. Some participants may not be comfortable disagreeing with the rest of the group or might be shy to speak their minds. Accordingly, UX researchers won't be able to uncover valuable user insights. In addition, there is a potential observer bias in interviews and focus groups, as the moderators can unintentionally influence the conversations by their reactions and follow-up questions. Also, managing and scheduling participants take time, especially if researchers want to gather several participants together. Using techniques like diary studies to gather long-term self-reported users' experiences has challenges, particularly finding several interested participants and keeping them motivated to continue the study. We highlight strengths and weaknesses of the several research methods in Table \ref{tab:researchmethods} to find out how we can gather more real, up-to-date, and comprehensive information about users' perceptions, concerns, and behaviors.

Conducting user research has some difficulties, as users' needs, motivations, and feelings may be concealed beneath a layer of civility, privacy, and triviality. Hence, HCI researchers and designers leveraged participatory exercises to explore users' needs and opinions they couldn't mention previously. The participatory design approach integrates different perspectives into the design process that are genuinely user-centered, leading to innovative solutions and increasing the chance of successful outcomes \cite{kensing1998participatory}. The selection of participants should be from different types of stakeholders (e.g., end-users, designers, researchers, product developers, etc.) to cover all points of view and avoid confusion and misleading in the design process. Similarly, gathering various stakeholders is time-consuming. It should be noted that executing other research methods such as interviews along with the participatory design is recommended to ensure nothing is overlooked.

HCI research encompasses both technology and human-behavioral aspects. Research contributions in HCI intend to uncover insights about human behaviors and their relationships to technology. Social media platforms are progressively being integrated into people's everyday lives. People use social media platforms for different reasons, including delivering messages \cite{pindayi2017social}, collaborating with others on a common interest \cite{lambton2020unplatformed}, creating an online community \cite{kotut2020preparing}, marketing and advertising products \cite{appel2020future}, sharing and presenting their opinions \cite{oh2015motivations}, and so on. As a result, HCI researchers are interested in exploring rich data on a large scale in social media platforms. The insights that come out of social media analysis, including unpacking and observing naturally occurring discussions and interactions on media, can inform HCI scholars and designers about hidden layers of information about users' needs, motivations, feelings, and experiences and strengthen their knowledge about targeted users.

\begin{figure}[h]
  \centering
  \includegraphics[width=\linewidth/2]{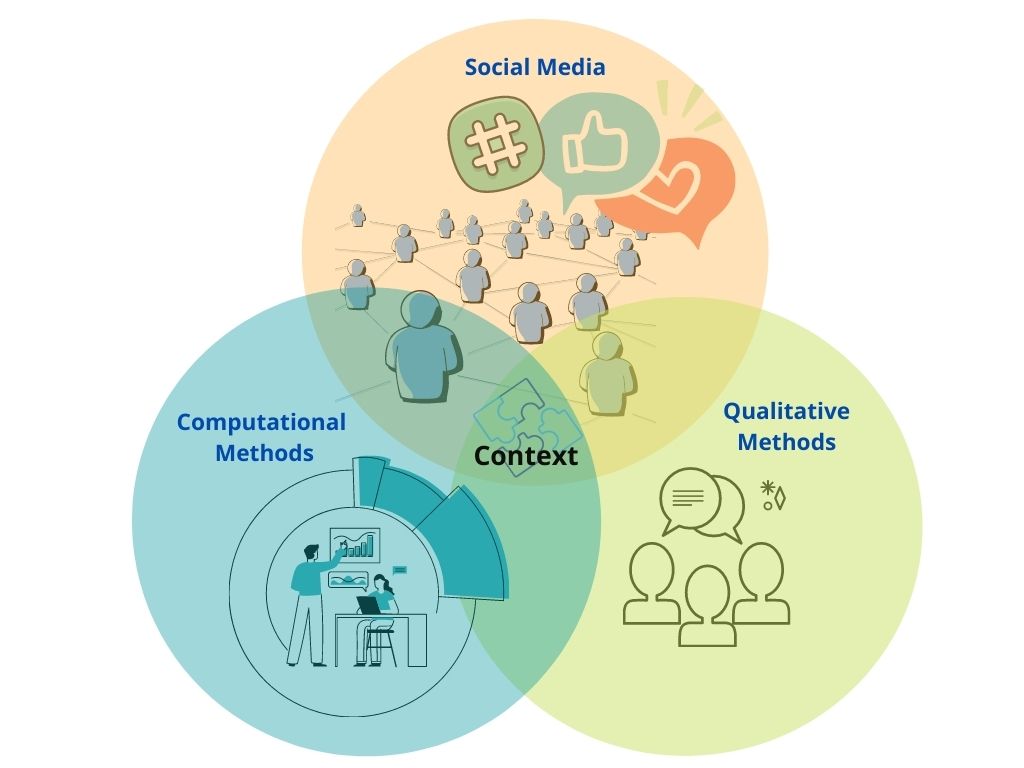}
  \caption{Showcasing the position of using social media along with traditional research methods}
  \label{Methodology}
\end{figure}

In exploring current HCI research practices with traditional methods, the shortcomings of those methods allows us to better see how social media technologies can improve, inform, and strengthen insights to better understand and represent user populations. Figure \ref{Methodology} shows how using social media platforms as supplementary tools along with traditional research methods and how social media platforms can be integrated to the design process. The following sections will provide some driving examples of leveraging social media through designing the research process and discusses the concerns linked with using and analyzing social media data. 

\begin{table*}
  \caption{Weaknesses and strengths of conducting some design research methods}
  \label{tab:researchmethods}
  \begin{tabular}{ccl}
    \toprule
    Research Method & Strengths & Weaknesses\\
    \midrule
    \texttt{Interviews} &  Being able to control the discussion/questions & Expensive and Time Consuming \\ & Observing non-verbal respondents' behaviors & Existence of subjective biases \\
    \\
    \hline
    \texttt{Focus Groups} & Group thinking and bringing out ideas & Strong opinions overshadow other group members \\ & Collecting qualitative and in-detail data & Difficulty in scheduling a time for getting together \\
    \\
    \hline
    \texttt{Diary Studies} & Understanding long-term and realistic behaviors & Sensitive to picking participants\\
    & Capturing external influencing factors & Needs to motivate participants to keep interested \\
    \\
    \hline
    \texttt{Social Media Analysis}& Collecting large-scale data across the world & Difficulty in analyzing big data in several platforms\\
    & Understanding users' unfocused expressions & Privacy issues \\
    \\
    \bottomrule
  \end{tabular}
\end{table*}

\section{Examples, Opportunities, and Concerns}

\subsection{I. Dig Deeper - Use of Social media for Interview Preparation}
Social media has been increasingly adopted in outdoor activities such as hiking and camping for planning and preparing, communicating on the trail, and sharing individual experiences \cite{kotut2020preparing}. In our ongoing study of technology for sustainable trail management on the Appalachian Trail (AT), a preliminary social media analysis has helped identify various themes of hikers' discussion and understand their priorities and concerns around sustainable practices. As we conduct interviews with trail managers, maintainers, and volunteer clubs along the AT, the insights from social media analysis have guided our discussions. These interactions have revealed the adoption of social media platforms by trail maintainers and managers for communicating with hikers and have enabled us to understand the social media practices of these organizations. 

However, as HCI researchers, we believe that users' social media presence offers a closer view of their verbal accounts. For instance, a trail club manager said, \textit{"we are doing better than other clubs as our Facebook page has the highest likes...We communicate the Leave No Trace principles when we find appropriate context."} Further information about the type of posts and engagement with users has been sought in the interviews, but such accounts remain unverified. Moreover, there are possibilities that these interviews provide little new information than that already present on social media. With prior information on activities of the interview subjects on social media, we can better frame the questions to gather deeper insights as well as record explanations for any discrepancies in the interviewers' responses and their social media representations.

\subsection{II. Sources - Use of Social media for validating Interviews}
As a rich source of information, social media has been leveraged in a criminal investigation as well by studying the digital trail of social media accounts \cite{lee2021}. A recent Netflix drama "Inventing Anna" \cite{anna2022} portrays the use of social media platform, Instagram, by the lead journalist in uncovering the facts about the unusual activities of the protagonist in New York City. With no leads for investigation, Instagram was used by the journalist to identify people as the potential sources of information and conduct interviews with them. These sources denied any relations with the protagonist. However, their posts on Instagram depicted otherwise. These posts were used as pieces of evidence of the protagonist' connections, enabling the journalist to trace them and collect information. Moreover, the posts' content, time, location, and event established the context for the interviews with the identified sources and helped point out any misrepresentations in the interview responses. So, as part of the design toolkit, social media analysis can substantiate the information collected through qualitative methods, like interviews, by uncovering and validating it. 


\subsection{III. Privacy}
Despite all the advantages of using social media data, the novelty of social media introduces concerns that need to be considered. Bringing up these concerns in conducting design and research methods will open discussion about leveraging the medium for futuring the user experience and interaction design. First, the privacy concerns with accessing and using social media data of the research subjects. While public profiles— of individuals and organizations— offer an open gateway to the data, it doesn't mean there aren't any risks associated with its use \cite{proferes2021studying}. Besides, private accounts require access permissions and more precautions to use their data. Ways to handle personal data (e.g., minimizing data, anonymizing data) \cite{di2021address} and include user consent into the design process require further exploration.

\subsection{IV. Large data concerns}
Another challenge with social media data is its massiveness, primarily stemming from the plethora of platforms available to users. Identifying the platform(s) to examine the digital footprint of research subjects is necessary. Users welcome every social media platform for a specific reason, as they represent different styles of engagement. For instance, Reddit is a social information aggregation platform that gathers users in communities, or subreddits on different topics. While Twitter and Instagram are both lower-effort platforms that support quick posts of text snippets and images, respectively. So, defining the criteria for the selection of relevant social media platforms is an important area for discussion to place social media into the design toolkit.

Furthermore, with varied usage practices and handles on multiple sites, gauging the need of synthesizing data from several social media sources at one place is open to debate. Organizations might post similar content on different platforms to reach wider audiences, while individual users' posts could be guided by the type of platform. For example, hikers are observed to use Reddit for pre-hike preparation and planning \cite{kotut2020preparing} and Instagram to share during- and post-hike experiences. Centralizing the information in the case of organizations is redundant while it could be useful in the case of individuals, although the scale of data processing remains questionable. 

\section{Conclusion and Future Work}
Social media has emerged as a force that shapes our interactions with technology, so it is important to integrate insights from it in the design process. We believe social media is a powerful resource to support HCI design and research in the age of remote life, as people use these platforms to keep in touch, update their information, and share their stories/experiences. Although HCI researchers and designers have been using social media to help them in various ways (e.g., preparing for conducting different research methods) and for different purposes (e.g., understanding online communities,  exploring users' opinions) \cite{snelson2016qualitative, kotut2020preparing, pometko2021drawing}, we believe that social media research/analysis should be considered more as a supplement design or research tool through tighter integration with traditional research methods (e.g., interviews, focus groups) to validate, enhance, and guide HCI-related studies. 
Therefore, a social media analysis tool in design research is worth exploring with stakeholders, practicing designers, researchers, students, and educators.

Our insights seek to inform HCI researchers and designers about the importance of using social media platforms during the design process and connecting them with traditional research methods, exploring related concerns such as privacy, and finding out proper ways to analyze big data to better represent the user populations. We believe that unpacking the users' perceptions is vital to designing a successful product. Hence, being aware of trend interests and values among users on social media will help UX designers and researchers deliver more pleasant user experiences to targeted users. We like to get feedback and gather new perspectives from prominent and knowledgeable professionals regarding analyzing social media and bringing it into the design process by attending this workshop to improve our work for future studies.







\bibliographystyle{ACM-Reference-Format}
\bibliography{sample-base}


\end{document}